\documentclass[journal=jctcce,manuscript=article]{achemso}

\SectionNumbersOn

\usepackage[version=3]{mhchem} 

\usepackage{multirow}
\usepackage{listings}
\usepackage{xspace}
\usepackage{float}
\usepackage{booktabs}
\usepackage{braket}
\usepackage{natmove}
\usepackage{todonotes}
\usepackage[flushleft]{threeparttable}
\usepackage[labelfont=bf]{caption}
\usepackage{subcaption}
\usepackage{pgfplots}
\usepackage{multirow}
\usepackage{threeparttable,array,siunitx}
\usepackage{float}
\usepackage{comment,algorithmicx, algorithm}
\usepackage{algcompatible}

\graphicspath{{./}}

\newcommand{\bigO}[1]{\ensuremath{ \mathcal{O}\left( #1 \right) }}
\newcommand{\tpno}{\ensuremath{\tau_\text{PNO}}}

\title{Optimized pair natural orbitals for the coupled cluster methods}

\author{Marjory C. Clement}
\affiliation[VT]{Department of Chemistry, Virginia Tech, Blacksburg, VA 24061, USA}

\author{Jinmei Zhang}
\affiliation{Computational Research Division, Lawrence Berkeley National Laboratory, Berkeley, CA 94720}

\author{Cannada A. Lewis}
\affiliation[VT]{Department of Chemistry, Virginia Tech, Blacksburg, VA 24061, USA}

\author{Chao Yang}
\email{cyang@lbl.gov}
\affiliation{Computational Research Division, Lawrence Berkeley National Laboratory, Berkeley, CA 94720}

\author{Edward F. Valeev}
\affiliation[VT]{Department of Chemistry, Virginia Tech, Blacksburg, VA 24061, USA}
\email{efv@vt.edu}

\begin{document}

\begin{abstract}
We present the coupled-cluster singles and doubles method formulated in terms of truncated pair-natural orbitals (PNO) that are optimized to minimize the effect of truncation. Compared to the standard ground-state PNO coupled-cluster approaches, in which truncated PNOs derived from first-order M\o ller-Plesset (MP1) amplitudes are used to compress the CC wave operator, the iteratively-optimized PNOs (``iPNOs'') offer moderate improvement for small PNO ranks but rapidly increase their effectiveness for large PNO ranks. The error introduced by PNO truncation in the CCSD energy is reduced by orders of magnitude in the asymptotic regime, with an insignificant increase in PNO ranks. The effect of PNO optimization is particularly effective when combined with Neese's perturbative correction for the PNO incompleteness of the CCSD energy. The use of the perturbative correction in combination with the PNO optimization procedure seems to produce the most precise approximation to the canonical CCSD energies for small and large PNO ranks. For the standard benchmark set of noncovalent binding energies remarkable improvements with respect to standard PNO approach range from a factor of 3 with PNO truncation threshold $\tpno=10^{-6}$ (with the maximum PNO truncation error in the binding energy of only 0.1 kcal/mol) to more than 2 orders of magnitude with $\tpno=10^{-9}$.
\end{abstract}

\section{Introduction}
First-principles many-body electronic structure methods can predict
many molecular properties, including structure, spectra,
thermodynamic data, and chemical reactivities. 
In particular, the coupled-cluster method\cite{Bartlett:2007CC}
can approach near-experimental accuracy for the prediction of the heats of formation
of small molecules,\cite{Tajti2004} with the hope that for heavier elements and larger systems, theory
will actually become more accurate than experiment.
Unfortunately, application of even the simplest many-body methods, such as
the coupled-cluster singles and doubles method (CCSD),\cite{Purvis1982} is restricted to small systems due to
the \bigO{N^6} computational complexity of its standard LCAO (``linear combination of atomic orbitals'')
implementation. Coupled-cluster methods that are capable of quantitative energetics,
such as the coupled-cluster singles, doubles, and
perturbative triples method [CCSD(T)],\cite{Raghavachari:CCSD(T)}
have an even higher computational complexity [in this particular case, \bigO{N^7}]. Thus, significant recent effort 
has focused on reducing the complexity of many-body methods.

The complexity of many-body methods can be reduced from their na\"{i}ve
figures by several techniques. First, the use of (non-LCAO) numerical
representations, e.g. real-space/reciprocal-space grids, that permit
fast application of operators can be used to reduce scaling,\cite{Bischoff:2012fw,Bischoff:2013cx,Schafer:2017kq,Mardirossian:2018fk} albeit
at the cost of increasing the verbosity of the representation. In the
context of LCAO, complexity reduction calls for the use of low-rank
representations (e.g. orbital localization, density fitting, and
iterative subspace compression) and screening. Recently, by combining
such techniques, practical reduced scaling implementations of LCAO
coupled-cluster methods, capable of maintaining chemically-acceptable
precision and achieving low-order (sometimes, linear) scaling with the system
size, have been demonstrated\cite{Pinski:2015ii,Riplinger:2016dq,Pavosevic:2017kb,Schwilk:2017ut,Ma:2017ef,Ma:2018fl,Schmitz:2014do,Schmitz:2016bu}.

A key to recent advances has been the introduction of pair natural orbitals (PNOs)\cite{Neese:2009_130_114108,Neese2009}
and other closely related concepts for block-wise compression.\cite{Yang:2011_OSV,Yang:2012_OSV,Riplinger:2013_TriplesNOs}
Just as the natural spin orbitals are the optimal basis (in the sense of wave function norm) for the CI expansion of
a 2-electron system,\cite{Lowdin1955} so PNOs provide an efficient (albeit not optimal) basis for encoding pair blocks of a wave operator.
Although PNOs date back to the 60s and 70s and the work of Edmiston and
Krauss,\cite{Edmiston:1966_PNO,Edmiston:1968_PNO}
Meyer,\cite{Meyer:1971_PNO,Meyer:1973_PNO,Meyer:1975_PNO}
Ahlrichs,\cite{Ahlrichs1975} and others, their recent use was popularized by the work of
Neese and co-workers,\cite{Neese:2009_130_114108,Neese2009}
who showed that they can reduce the scaling and prefactor to a degree sufficient for
early crossover with canonical methods. A combination of PNO-style compression
with local formulations of coupled-cluster (already shown to be capable of linear scaling by Werner and co-workers\cite{Schutz:2001_LCCSD,Schutz:2002,Schutz:2003})
gives rise to reduced\cite{Schmitz:2014do,Schmitz:2016bu} and even linear scaling\cite{Riplinger:2013ek,Riplinger:2013tn,Riplinger:2016dq,Pavosevic:2016bc,Pavosevic:2017kb,Saitow:2017bo,Schwilk:2017ut,Ma:2017ef,Ma:2018fl} variants of the PNO coupled-cluster methods, which achieve practical supremacy compared
to the canonical coupled-cluster implementations for systems with 10-20 atoms.

The use of PNOs in any infinite-order method, such as configuration interaction, coupled-cluster, or Green's function approaches,
is predicated on access to guess two-body amplitudes of sufficient quality to construct accurate PNOs.
All modern applications use (approximate) first-order M\o ller-Plesset (MP1) amplitudes\cite{Neese:2009_130_114108}
to form the PNOs, although other choices have been investigated.\cite{Meyer:1971_PNO,Meyer:1973_PNO}
It is clear that such a choice may be suboptimal, such as for cases when correlation can introduce substantial relaxation effects (e.g. in anions)
and in small-gap systems in general (conjugated organic molecules, semiconductor crystals).
Here we propose to investigate how closely MP1-based PNOs approximate the {\em optimal} PNOs in the context of the coupled-cluster singles and doubles method.
To construct optimal PNOs we have devised an iterative algorithm for refinement of the PNOs; the moniker ``iPNO'' will be used to distinguish
these optimized PNOs from standard PNOs.


Our paper is organized as follows. In Section~\ref{background},
we outline the construction and truncation of PNOs.
We then discuss the current method for iPNO construction in more detail and
include a discussion of perturbative corrections for the PNO truncation errors that we investigated in this work.
Section~\ref{results} details numerical performance of iPNO-CCSD vs standard PNO-CCSD and canonical CCSD.
We summarize our findings and discuss potential for the use of iPNOs in
a production-quality PNO coupled-cluster implementation in Section~\ref{conclusions}.

\section{\label{background}Theoretical background}

The pair natural orbitals of pair $ij$ are the eigenvectors of the corresponding pair density $\mathbf {D}^{ij}$:
\begin{align}
\label{eq:Uij}
\mathbf{D}^{ij} \mathbf{U}^{ij} = \mathbf{U}^{ij} \mathbf{n}^{ij},
\end{align}
where $(\mathbf{U}^{ij})_{b a} \equiv U_{b a_{ij}}$ is the $b$th expansion coefficient of PNO $a_{ij}$,
and $(\mathbf{n}^{ij})_{ab} \equiv n_{a_{ij}} \delta_{a_{ij} b_{ij}}$ is the associated PNO occupation number.\footnote{Following convention, we have used $i$, $j$, \dots;  $a$, $b$, \dots; and $p$, $q$, \dots for the occupied, virtual, and general orbitals in the Hartree-Fock (HF) basis, respectively.}
The pair density matrix is defined by the two-body amplitudes $\mathbf{T}^{(ij)}$:
\begin{align}
 \label{eq:Dij}
\mathbf {D}^{ij} & = 
\frac{1}{1 + \delta_{ij}} 
\left(\tilde{\mathbf{T}}^{ij \dagger} \mathbf{T}^{ij} + 
\tilde{\mathbf{T}}^{ij} \mathbf{T}^{ij \dagger}\right),
\end{align}
where $(\mathbf{T}^{ij})_{ab} \equiv T^{ij}_{ab}$ and $\tilde{T}^{ij}_{ab} = 2 T^{ij}_{ab} - T^{ij}_{ba}$.
Transforming amplitudes to the full set of PNOs for each pair,
\begin{align}
\bar{\mathbf{T}}^{ij} \equiv \mathbf{U}^{ij} \mathbf{T}^{ij} \mathbf{U}^{ij \dagger},
\end{align}
does not produce any computational savings but, rather, greatly increases the costs of computing Hamiltonian matrix elements in the LCAO representation, e.g. the order-4 tensor $g^{ab}_{cd} \equiv \bra{cd} r_{12}^{-1} \ket{ab}$, traditionally computed at an \bigO{N^5} cost, becomes, in the PNO basis, $g^{a_{ij} b_{ij}}_{c_{ij} d_{ij}}$, which requires an \bigO{N^7} effort to compute.
Computational savings are realizable if PNOs are truncated to include only those orbitals for which $n_{a_{ij}} \geq \tpno$, where truncation threshold $\tpno$ is a user-defined model parameter (setting $\tpno=0$ makes the PNO-based representation exact). For any finite $\tpno$ the number of PNOs per pair is independent of the system size for systems with nonzero gap, so parametrizing the wave operator in terms of the PNO basis amplitudes $\left(\bar{\mathbf{T}}^{ij}\right)_{a_{ij}b_{ij}}$ directly reduces the number of parameters from
\bigO{N^4} to \bigO{N^2}. One-body amplitudes $\bar{\mathbf{T}}^{i}$ are similarly expressed in the compressed basis of orbital-specific virtuals (OSVs) $\mathbf{U}^{i}$, here obtained for orbital $i$ as the PNOs of pair $ii$ truncated with threshold $\tpno/100$; this effectively makes the effect of OSV truncation completely negligible relative to that of PNO
truncation.

The precision of the PNO representation is determined by the truncation parameter $\tpno$ {\em and} the quality of the guess amplitudes used to 
compute the PNOs. Although a variety of types of guess amplitudes have been used to construct PNOs in the past,\cite{Edmiston:1966_PNO,Edmiston:1968_PNO,Ahlrichs1975,Meyer:1971_PNO,Meyer:1973_PNO,Meyer:1975_PNO} in recent work,
the PNOs are usually computed from exact or approximate first-order M\o ller-Plesset (MP1) amplitudes,\cite{Neese:2009_130_114108}
which, in the basis of canonical Hartree-Fock orbitals, are computed as
\begin{align}
\label{eq:T_mp1}
 T^{ij}_{ab} = 
 \frac{g^{ij}_{ab}}{f^i_i + f^j_j - f^a_a - f^b_b},
\end{align}
where $f^p_q =  \bra{p} \hat{f} \ket{q}$ are the matrix elements of the Fock operator. If the occupied orbitals are localized, the amplitudes
evaluated via Eq. \eqref{eq:T_mp1} are referred to as semicanonical amplitudes.\cite{Neese2009} While semicanonical amplitudes
are not the exact MP1 amplitudes, they are sufficiently accurate for the purpose of computing PNOs.
This approach has also been generalized by Tew and co-workers to the context of explicitly correlated methods.\cite{Tew:2011}

Here we propose to explore whether it is possible to improve MP1 PNOs in the context of iterative
solvers like those in the coupled-cluster method. The idea is to update PNOs periodically
using the current values of the CC doubles amplitudes. In cases where the correlation effects are
not described well by perturbation theory and the MP1 amplitudes are a poor approximation
to the exact CC doubles amplitudes, updating PNOs might produce substantial savings and/or higher accuracy at
constant compression rank.

Since the definition of the pair density in  Eq. \eqref{eq:Dij} includes
the amplitudes expressed in the full space of unoccupied orbitals, it would appear that
updating PNOs is only possible if guess amplitudes can be periodically computed in the full space
by e.g. computing the residuals of the CC amplitude equations in the full space also.
As we discuss later, it should be possible to update PNOs without ever constructing $\mathbf{T}^{ij}$ in the full space of unoccupied orbital products.
Since our goal here is to assess the performance of the PNOs optimized for the coupled-cluster
family of methods,
we utilize a canonical CCSD solver rather than a production PNO CCSD solver (preliminary testing
of these ideas utilized a pilot PNO-CCD solver). This {\em simulated} implementation is an appropriate choice for testing
the approach since coupled-cluster residuals in the full space of unoccupied states
are directly available. Note that simulation of a PNO CC solver using a canonical CC solver has been utilized before by Werner and co-workers\cite{Korona:2003_simulation,Krause:2012_simulation}
and recently by us in the context of PNO-EOM-CCSD.\cite{Peng:2018_PNO-EOM-CCSD}

The iPNO-CCSD solver is described in Algorithm~\ref{alg:iPNOCCSD}.
The basic idea is to solve the CCSD amplitude equations in a given fixed PNO subspace (we refer to these iterations as microiterations)
and iteratively update the subspace by reconstructing the amplitudes in the full space and recomputing the PNOs (these are macroiterations).

\begin{algorithm}
\begin{center}
  \begin{minipage}{6in}
\begin{tabular}{p{0.5in}p{5.5in}}
{\bf Input}:  &  \begin{minipage}[t]{5in}
             micro iteration convergence predicate $C_\text{micro}(\mathbf{R}^{ij}, \mathbf{R}^{i})$;\\
             macro iteration convergence predicate $C_\text{macro}(\mathbf{R}^{ij}, \mathbf{R}^{i})$;\\
             PNO truncation threshold \tpno.\\
                  \end{minipage} \\
{\bf Output}:  &  \begin{minipage}[t]{5in}
             Converged doubles $\{\mathbf{T}^{ij}\}$
             and singles $\{\mathbf{T}^{i}\}$ CCSD amplitudes expressed in the basis of optimized PNOs $\mathbf{U}^{ij}$ and OSVs $\mathbf{U}^{i}$; \\  
                  \end{minipage}
\end{tabular}
\begin{algorithmic}[1]
\STATE $k = 0$;
\STATE $\{\mathbf{T}^{i}_{(0)}\} = 0$;
\STATE Initialize $\{ \mathbf{T}^{ij}_{(0)} \}$ to semicanonical MP1 amplitudes using Eq. \eqref{eq:T_mp1};
\STATE Construct pair densities $\{\mathbf{D}^{ij}\}$ as well as truncated PNOs, $\{\mathbf{U}^{ij}\}$, and OSVs, $\{\mathbf{U}^{i}\}$ using Eqs. \eqref{eq:Dij} and \eqref{eq:Uij};
\REPEAT \label{step:macro}
  \REPEAT \label{step:micro}
    \STATE Compute CCSD residuals $\{\mathbf{R}^{ij}\}$ and $\{\mathbf{R}^{i}\}$ using amplitudes $\{ \mathbf{T}^{ij}_{(k)} \}$ and $\{ \mathbf{T}^{i}_{(k)} \}$ ;
    \STATE Transform 2-body residual to PNO basis: $\bar{\mathbf{R}}^{ij} = \mathbf{U}^{ij \dagger} \mathbf{R}^{ij} \mathbf{U}^{ij}, \forall ij$ ;
    \STATE Transform 1-body residual to OSV basis: $\bar{\mathbf{R}}^{i} = \mathbf{U}^{i \dagger} \mathbf{R}^{i} \mathbf{U}^{i}, \forall i$ ;
    \STATE Use $\{\bar{\mathbf{R}}^{ij}\}$ and $\{\bar{\mathbf{R}}^{i}\}$ to compute Jacobi/DIIS updates for 2-body PNO basis amplitudes, $\{ \bar{\mathbf{\Delta}}^{ij} \}$, and
              1-body OSV basis amplitudes, $\{ \bar{\mathbf{\Delta}}^{i} \}$ ; 
       \STATE Back transform $\bar{\mathbf{\Delta}}^{ij}$ and
              $\bar{\mathbf{\Delta}}^{i}$ to the full unoccupied
              space: \\
              $\bar{\mathbf{\Delta}}^{ij} =  \mathbf{U}^{ij} \mathbf{\Delta}^{ij} \mathbf{U}^{ij  \dagger}, \forall ij$, \\
              $\bar{\mathbf{\Delta}}^{i} =  \mathbf{U}^{i} \mathbf{\Delta}^{i} \mathbf{U}^{i  \dagger}, \forall i$; \\
       \STATE Update the amplitudes:\\
              $\mathbf{T}^{ij}_{(k+1)} = \mathbf{T}^{ij}_{(k)} + \mathbf{\Delta}^{ij}, \forall ij$,\\
              $\mathbf{T}^{i}_{(k+1)} = \mathbf{T}^{i}_{(k)} + \mathbf{\Delta}^{i}, \forall ij$;
       \STATE $k\leftarrow k+1$;
  \UNTIL {not $C_\text{micro}(\{\mathbf{R}^{ij}\},\{\mathbf{R}^{i}\})$ }
  
  \STATE Compute CCSD residuals $\{\mathbf{R}^{ij}\}$ and $\{\mathbf{R}^{i}\}$ using amplitudes $\{ \mathbf{T}^{ij}_{(k)} \}$ and $\{ \mathbf{T}^{i}_{(k)} \}$ ;
  \STATE Use $\{\mathbf{R}^{ij}\}$ and $\{\mathbf{R}^{i}\}$ to compute Jacobi/DIIS updates for 2-body amplitudes, $\{ \mathbf{\Delta}^{ij} \}$, and
              1-body amplitudes, $\{ \mathbf{\Delta}^{i} \}$ ; 
       \STATE Compute amplitudes for the PNO update:\\
              $\mathbf{\hat{T}}^{ij} = \mathbf{T}^{ij}_{(k)} + \mathbf{\Delta}^{ij}, \forall ij$,\\
              $\mathbf{\hat{T}}^{i} = \mathbf{T}^{i}_{(k)} + \mathbf{\Delta}^{i}, \forall i$;
      \STATE Construct updated truncated PNOs, $\{\mathbf{U}^{ij}\}$, and OSVs, $\{\mathbf{U}^{i}\}$, using $\{\mathbf{\hat{T}}^{ij}\}$;
      \STATE Project amplitudes $\{\mathbf{T}^{ij}_{(k)}\}$ and $\{\mathbf{T}^{i}_{(k)}\}$ to the updated PNO and OSV subspaces, respectively:\\
             $\mathbf{T}^{ij}_{(k+1)} \leftarrow \mathbf{U}^{ij} \mathbf{U}^{ij \dagger}\mathbf{T}^{ij}_{(k)} \mathbf{U}^{ij} \mathbf{U}^{ij \dagger}, \forall ij,$\\
             $\mathbf{T}^{i}_{(k+1)} \leftarrow \mathbf{U}^{i} \mathbf{U}^{i \dagger}\mathbf{T}^{i}_{(k)} \mathbf{U}^{i} \mathbf{U}^{i \dagger}, \forall i;$
      \STATE $k\leftarrow k+1$;
\UNTIL {not $C_\text{macro}(\{\mathbf{R}^{ij}\},\{\mathbf{R}^{i}\})$ }
\end{algorithmic}
\end{minipage}
\end{center}
\caption{iPNO-CCSD simulation algorithm\label{alg:iPNOCCSD}}
\end{algorithm}

\section{Computational details}
The iPNO-CCSD approach was implemented in a developmental version of the
Massively Parallel Quantum Chemistry package (version 4).\cite{MPQC4}
Initial assessment of the iPNO-CCSD approach utilized a representative 12-system subset (see Table~\ref{systems})
of the S66 data set,\cite{Hobza:2011_S66} the geometries of which were
taken from the Benchmark Energy and Geometry Database (BEGDB).\cite{begdb}
The full S66 data set was used in the final comparison of iPNO-CCSD with standard PNO-CCSD.
For all calculations, the cc-pVDZ-F12 basis set\cite{ccpVDZF12}
was employed, with all two-electron integrals approximated by density fitting
in the aug-cc-pVDZ-RI basis set\cite{augccpVDZRI}.

\begin{table}
\begin{center}
\begin{tabular}{ccc}
\hline\hline
\textbf{Index} & \textbf{Index in S66} & \textbf{System} \\
\hline
 1 & 1 & Water \dots Water \\
 2 & 2 & Water \dots MeOH \\
 3 & 3 & Water \dots $\text{MeNH}_2$ \\
 4 & 9 & $\text{MeNH}_2$ \dots MeOH \\
 5 & 24 & Benzene \dots Benzene ($\pi$-$\pi$) \\
 6 & 25 & Pyridine \dots Pyridine ($\pi$-$\pi$) \\
 7 & 26 & Uracil \dots Uracil ($\pi$-$\pi$) \\
 8 & 34 & Pentane \dots Pentane \\
 9 & 47 & Benzene \dots Benzene (TS) \\
 10 & 50 & Benzene \dots Ethyne (CH-$\pi$) \\
 11 & 59 & Ethyne \dots Water (CH-O) \\
 12 & 66 & $\text{MeNH}_2$ \dots Pyridine \\
\hline\hline
\end{tabular}
\caption{The 12-system subset of the S66 benchmark dataset utilized for initial assessment of the iPNO-CCSD approach.}
\label{systems}
\end{center}
\end{table}

\section{\label{results}Results and discussion}

\subsection{PNO-CCSD vs. iPNO-CCSD}

\begin{figure}[h]
\centering
  \begin{subfigure}{0.49\textwidth}
    \centering
    \includegraphics[width=\textwidth]{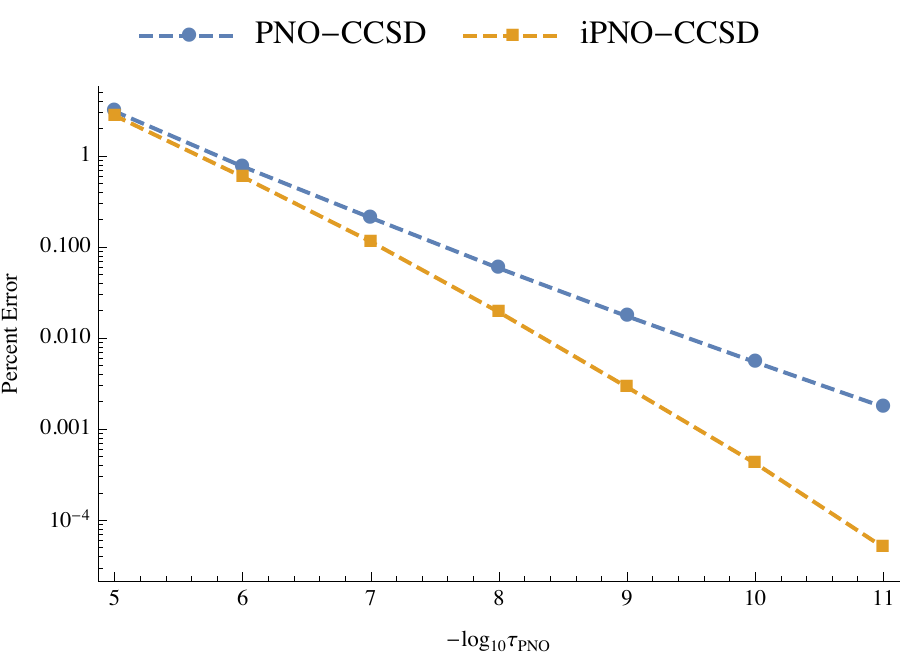}
    \caption{Maximum absolute error (MAX)}
    \label{fig:uncorrMAXdim}
  \end{subfigure}
  \hfill
  \begin{subfigure}{0.49\textwidth}
    \centering
    \includegraphics[width=\textwidth]{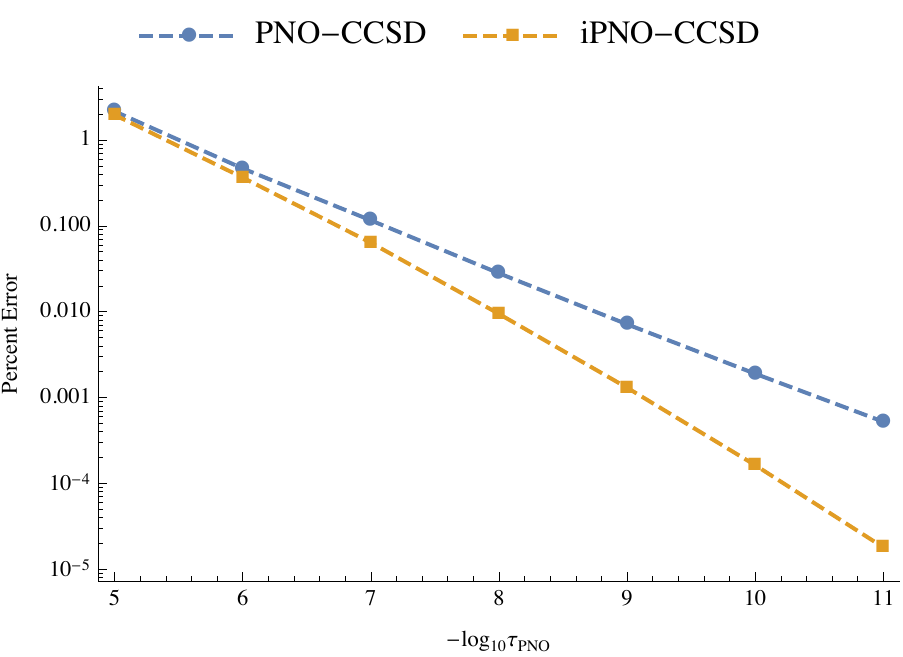}
    \caption{Mean absolute error (MAE)}
    \label{fig:uncorrMAEdim}
  \end{subfigure}
  \caption{Statistical analysis of dimer absolute correlation energy percent errors}
  \label{fig:dimerStats}
\end{figure}

\begin{figure}[h]
\centering
  \includegraphics[width=0.5\textwidth]{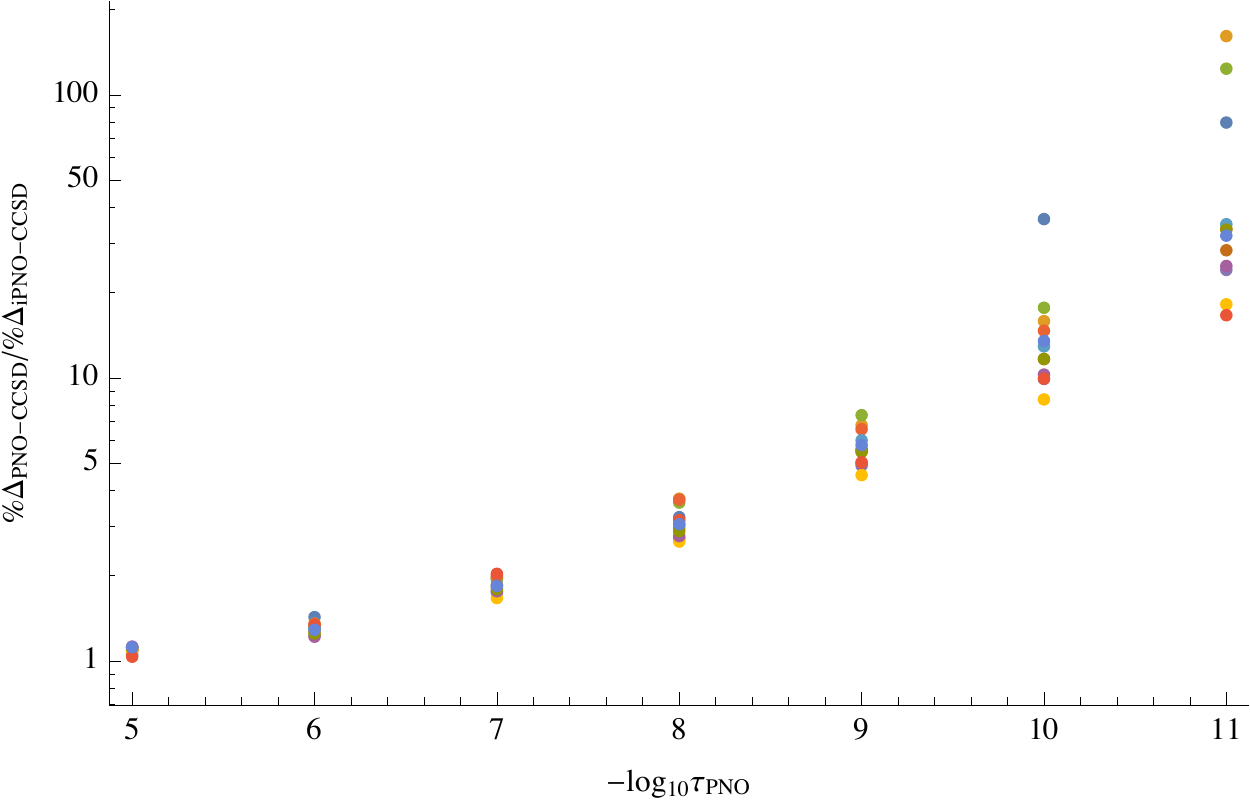}
  \caption{Ratio of PNO-CCSD dimer correlation energy percent error to iPNO-CCSD dimer correlation energy percent error for all systems}
  \label{fig:PerErrortPNOratio}
\end{figure}

First we compare the performance of the PNO and iPNO methods for the absolute correlation energies of dimers.
Figure~\ref{fig:dimerStats} illustrates the max and mean absolute percent error in dimer absolute energy, relative to the canonical CCSD energy, as a function of $\tpno$.
It is clear that the iPNO-CCSD approach performs consistently better than the PNO-CCSD
scheme; a modest average improvement of a factor of $1.9$ at $\tpno=10^{-7}$ (the value used in routine application of PNO methods)
becomes an improvement of more than an order of magnitude for $\tpno<10^{-10}$.
Figure~\ref{fig:PerErrortPNOratio} illustrates this improvement in more detail.

\begin{figure}[h]
\centering
  \includegraphics[width=\textwidth]{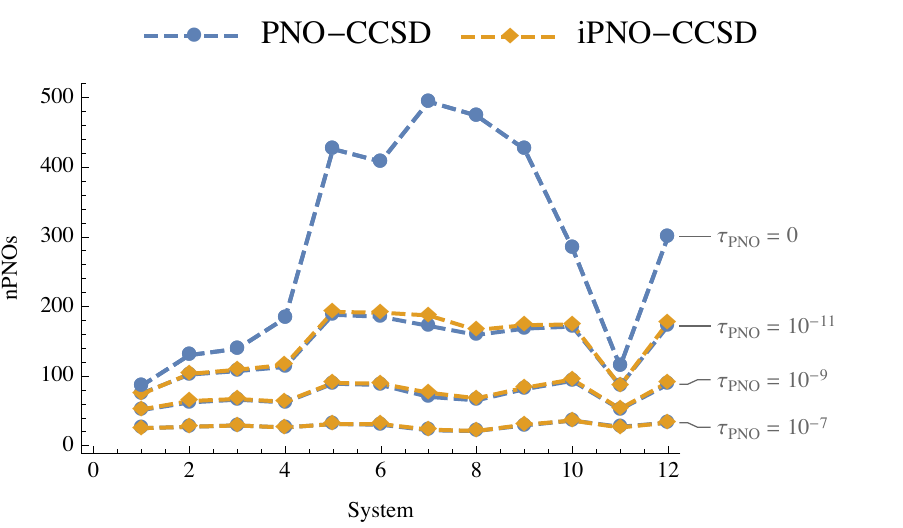}
  \caption{The average number of PNOs/pair for three \tpno values, compared across
  all three methods}
  \label{fig:rankPlot}
\end{figure}

The smaller truncation errors of the iPNO-CCSD correlation energies relative to their standard PNO-CCSD counterparts do not come at the cost of increased PNO ranks, as is demonstrated in Figure~\ref{fig:rankPlot}.

\begin{figure}[h]
\centering
  \begin{subfigure}{0.49\textwidth}
    \centering
    \includegraphics[width=\textwidth]{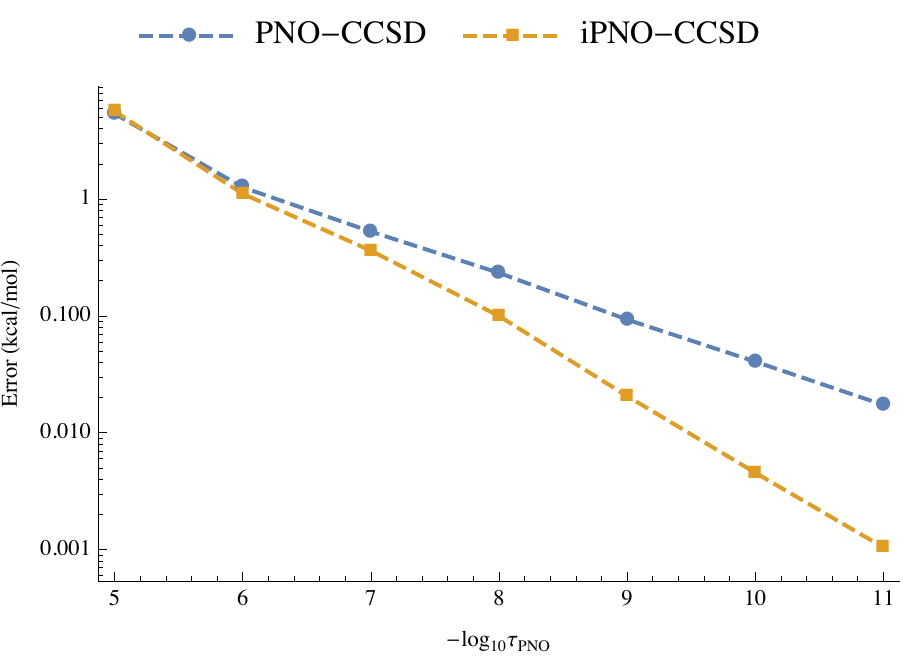}
    \caption{Maximum absolute error (MAX)}
    \label{fig:uncorrMAXbe}
  \end{subfigure}
  \hfill
  \begin{subfigure}{0.49\textwidth}
    \centering
    \includegraphics[width=\textwidth]{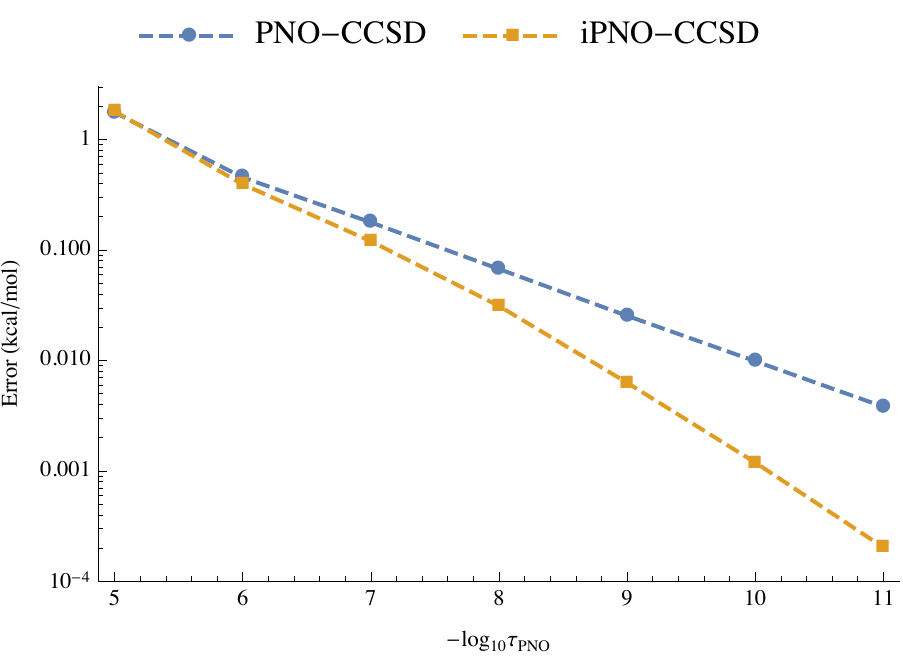}
    \caption{Mean absolute error (MAE)}
    \label{fig:uncorrMAEbe}
  \end{subfigure}
  \caption{Statistical analysis of binding energy absolute errors}
  \label{fig:BEstats}
\end{figure}

Of course, in chemistry we are usually interested in differences of correlation energies.
The performance of iPNO-CCSD for the binding energies of the dimers studied
is compared to that of PNO-CCSD in Figure~\ref{fig:BEstats}.
It appears that the improved performance of iPNO-CCSD for the absolute correlation energies translates into improved performance for the binding energies as well.

\subsection{Perturbative Energy Correction for PNO Incompletness}
As the preceeding data indicate, PNO optimization decreases the PNO truncation error in the CCSD energy; in other words, for the same rank,
the iPNO-CCSD energy is closer to the canonical CCSD energy than its standard PNO-CCSD counterpart. An interesting follow-up
question is whether the reduction in the PNO truncation error can be achieved in another way. Neese and co-workers proposed
a perturbative correction for the PNO truncation,\cite{Neese:2009_130_114108,Neese2009} obtained as the difference between the
(semicanonical) MP2 and PNO-MP2 energies, both easily available in the course of computing MP1 PNOs:
\begin{align}
\Delta_\text{PNO-MP2} \equiv & E_\text{MP2} - E_\text{PNO-MP2},  \quad \text{where} \\
E_{\text{MP}2} = & \sum_{\substack{ij\\ab}}
\left(2g^{ij}_{ab}-g^{ij}_{ba}\right)t^{ij}_{ab}, \\
E_{\text{PNO-MP}2} = & \sum_{\substack{ij\\\bar{a}_{ij}\bar{b}_{ij}}}
\left(2g^{ij}_{\bar{a}_{ij}\bar{b}_{ij}}-g^{ij}_{\bar{b}_{ij}\bar{a}_{ij}}\right)
t^{ij}_{\bar{a}_{ij}\bar{b}_{ij}},
\end{align}
and where both standard and PNO unoccupied orbitals are assumed to be canonical, i.e. the Fock operator is diagonal in these spaces.
An improved estimate of the canonical CCSD energy is then obtained as
\begin{equation}
E_{\text{CCSD}} \approx E_{\text{PNO-CCSD}} + \Delta_{\text{PNO-MP2}} .
\end{equation}
If the PNO incompleteness errors of the PNO-MP2 and PNO-CCSD energies were identical, this correction would be exact;
thus the key assumption of this scheme is that the PNO incompleteness is not sensitive to the level of correlation treatment.
It seems to be a good assumption in practice: Neese and co-workers observed\cite{Neese:2009_130_114108,Neese2009}
that, for the practical values of $\tpno$, the use of the perturbative correction significantly reduces the PNO incompleteness error.

Thus we decided to investigate whether the observed reduction in the PNO incompleteness of the correlation energy
due to the optimization of PNOs is accounted for by the $\Delta_{\text{PNO-MP2}}$ correction.
Since the change in the PNO basis should be accommodated by the incompleteness correction,
by analogy with $\Delta_\text{PNO-MP2}$, we proposed the use of the following correction for iPNO-CCSD energies:
\begin{align}
\Delta_\text{iPNO-MP2} \equiv & E_\text{MP2} - E_\text{iPNO-MP2},
\end{align}
where $E_\text{iPNO-MP2}$ is evaluated exactly as $E_\text{PNO-MP2}$ but using optimized CCSD PNOs as the basis.
All quantities needed to compute semicanonical $E_\text{PNO-MP2}$ and $E_\text{iPNO-MP2}$ are
readily available in the iPNO-CCSD code, and the implementation is straightforward.

\begin{figure}[h]
\centering
  \begin{subfigure}{0.49\textwidth}
    \centering
    \includegraphics[width=\textwidth]{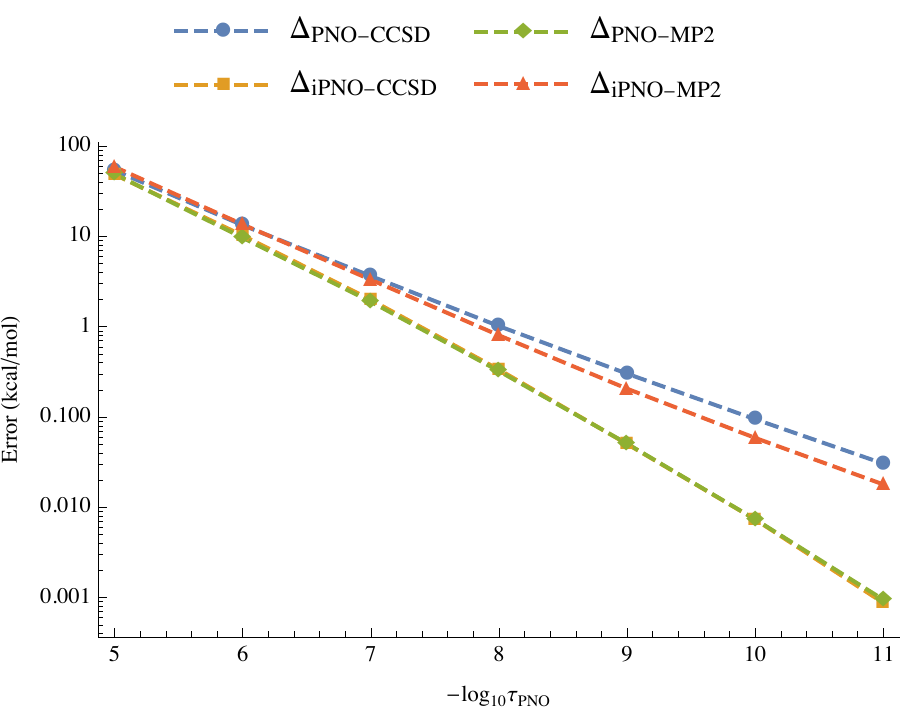}
    \caption{Maximum absolute error (MAX)}
    \label{fig:dimerDeltaMAX}
  \end{subfigure}
  \hfill
  \begin{subfigure}{0.49\textwidth}
    \centering
    \includegraphics[width=\textwidth]{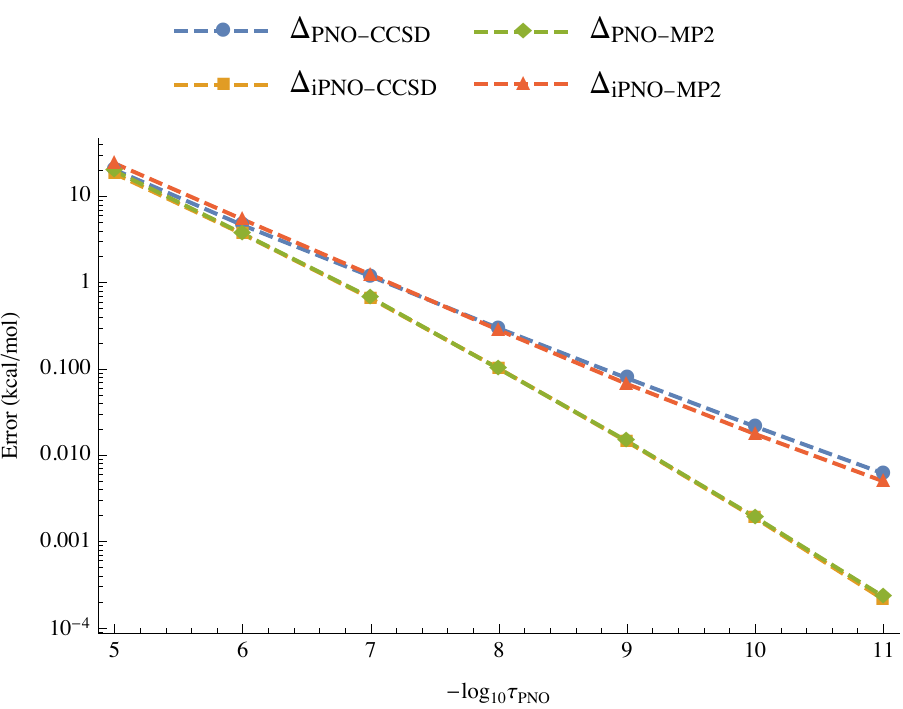}
    \caption{Mean absolute error (MAE)}
    \label{fig:dimerDeltaMAE}
  \end{subfigure}
  \caption{Statistical analysis of dimer energy absolute errors and perturbative
  energy corrections}
  \label{fig:dimerDeltaStats}
\end{figure}

\begin{figure}[h]
\centering
  \begin{subfigure}{0.49\textwidth}
    \centering
    \includegraphics[width=\textwidth]{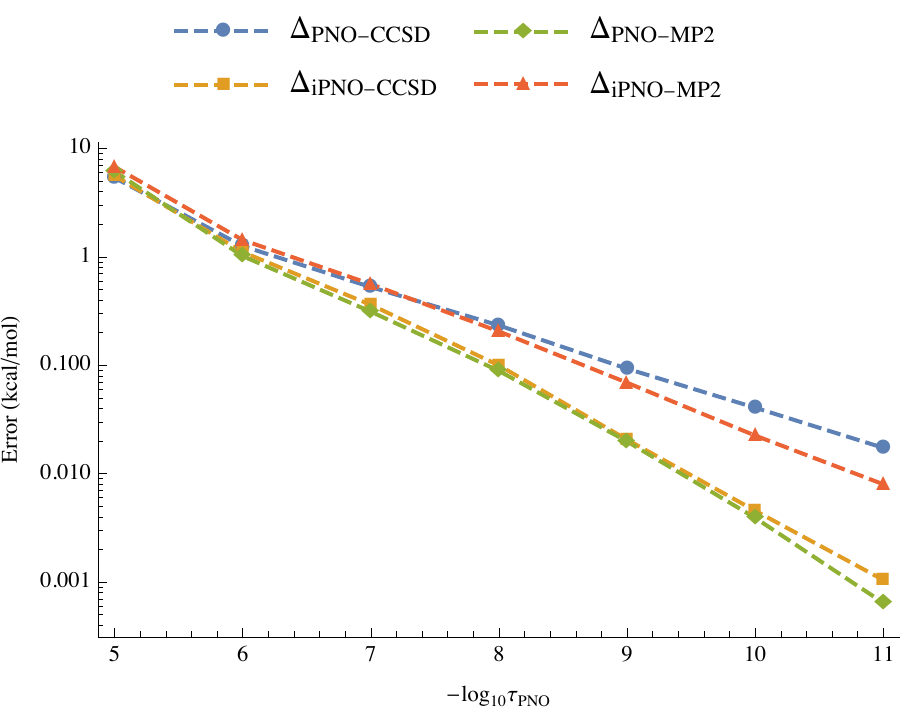}
    \caption{Maximum absolute error (MAX)}
    \label{fig:beDeltaMAX}
  \end{subfigure}
  \hfill
  \begin{subfigure}{0.49\textwidth}
    \centering
    \includegraphics[width=\textwidth]{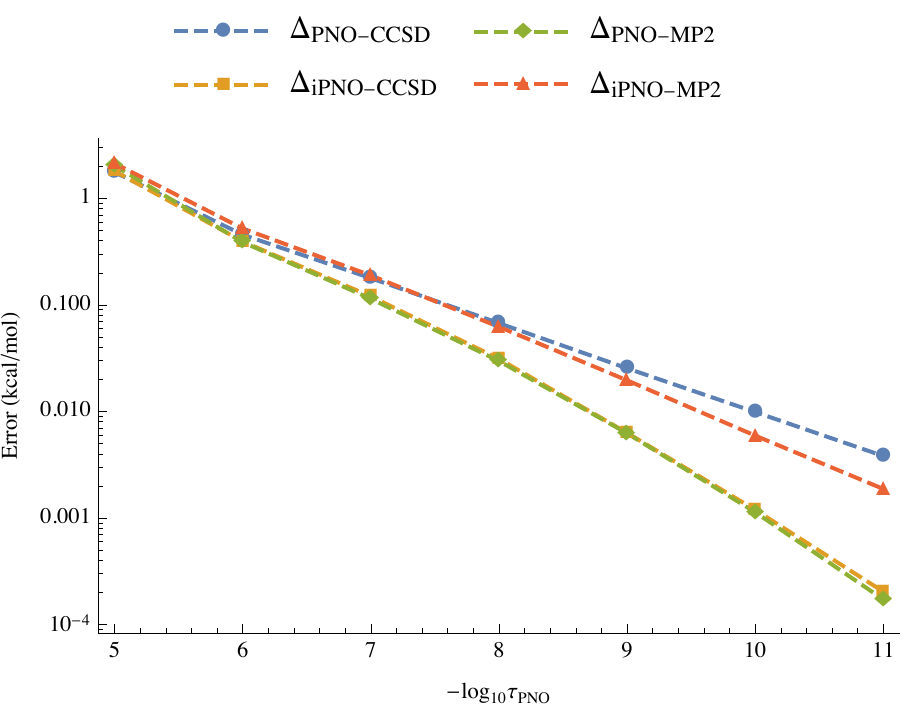}
    \caption{Mean absolute error (MAE)}
    \label{fig:beDeltaMAE}
  \end{subfigure}
  \caption{Statistical analysis of binding energy absolute errors and perturbative
  energy corrections}
  \label{fig:beDeltaStats}
\end{figure}

The max and mean absolute errors of the PNO-CCSD and iPNO-CCSD correlation energies of the dimers relative to their canonical CCSD counterparts
are compared to the $\Delta_{\text{PNO-MP2}}$ and $\Delta_{\text{iPNO-MP2}}$ corrections in Figure~\ref{fig:dimerDeltaStats}, with the corresponding data for binding energies
shown in Figure~\ref{fig:beDeltaStats}.
It is clear that, with coarsely truncated PNOs (large $\tpno$), $\Delta_{\text{PNO-MP2}}$ is nearly indistinguishable from the error in PNO-CCSD,
and hence it effectively corrects for the PNO incompleteness error, as observed by Neese {\em et al.} However, as $\tpno$ decreases, the quality of the correction decreases
and already at $\tpno = 10^{-8}$, $\Delta_{\text{PNO-MP2}}$ becomes ineffective.

Another observation is that the proposed $\Delta_{\text{iPNO-MP2}}$ correction
is not effective for correcting iPNO-CCSD energies; for $\tpno< 10^{-7}$ $\Delta_{\text{iPNO-MP2}}$ significantly overcorrects $E_\text{iPNO-CCSD}$.

The most interesting suggestion drawn from the data in Figures~\ref{fig:dimerDeltaStats} and \ref{fig:beDeltaStats} is that $\Delta_{\text{PNO-MP2}}$ seems to be an ideal PNO truncation correction for
$E_\text{iPNO-CCSD}$ for all values of $\tpno$. This observation is seemingly counterintuitive, since $\Delta_{\text{PNO-MP2}}$ is computed using MP1 PNOs,
whereas $E_\text{iPNO-CCSD}$ uses relaxed CCSD PNOs. While an in-depth investigation of this effect is outside the scope of this work, a possible line of inquiry to explain
this observation goes as follows. The PNO truncation error in standard PNO-CCSD is driven by two effects: the error due to the use of suboptimal (i.e. MP1) PNOs and due to the
fixed rank error. It is clear that the use of MP1 PNOs in PNO-CCSD results in substantial errors due to the suboptimal PNOs; for high PNO ranks (small $\tpno$) the PNO truncation error of standard PNO-CCSD is entirely dominated by the suboptimality of PNOs, as evidenced by the massive reduction of the PNO incompleteness by PNO optimization (compare iPNO-CCSD and PNO-CCSD errors). The use of {\em optimal} PNOs, i.e. MP1 PNOs for PNO-MP2 and CCSD PNOs for iPNO-CCSD, minimizes the errors in respective correlation energies {\em for fixed PNO ranks} and thus eliminates one component of the error. It seems that the second contribution to the error, due to the fixed PNO rank, is almost identical
for MP2 and CCSD energies; further investigation of this phenomenon is left to future studies.

\begin{figure}[h]
\centering
  \begin{subfigure}{0.49\textwidth}
    \centering
    \includegraphics[width=\textwidth]{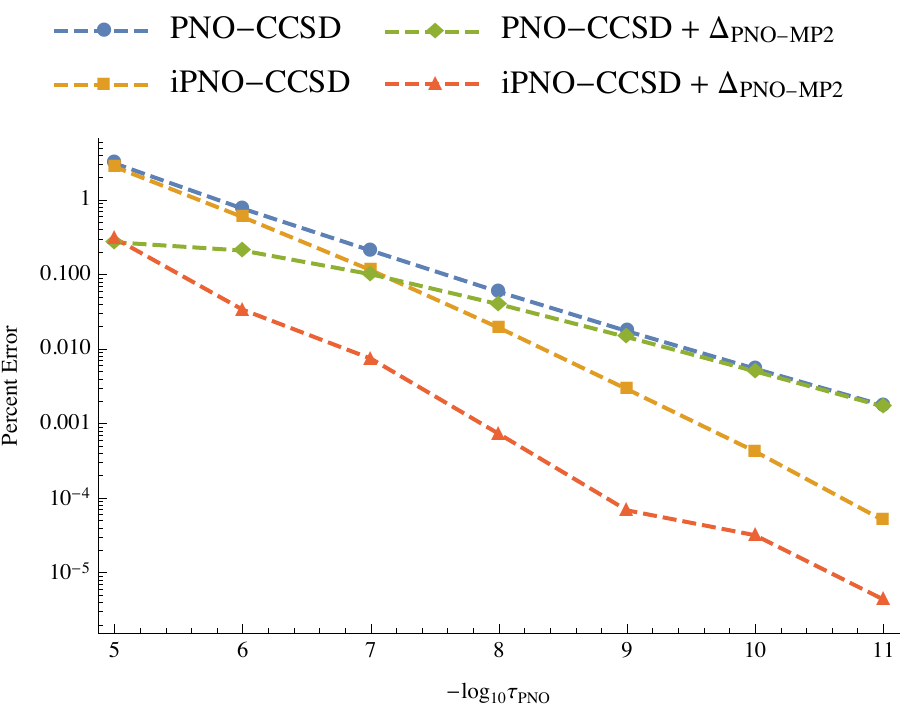}
    \caption{Maximum absolute error (MAX)}
    \label{fig:dimerCorUncorMAX}
  \end{subfigure}
  \hfill
  \begin{subfigure}{0.49\textwidth}
    \centering
    \includegraphics[width=\textwidth]{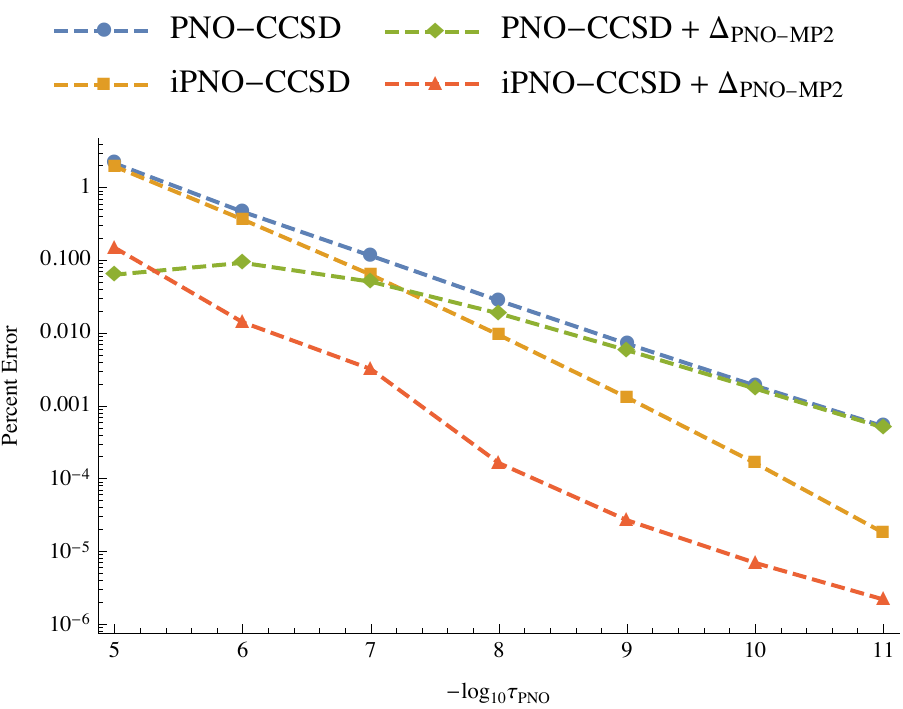}
    \caption{Mean absolute error (MAE)}
    \label{fig:dimerCorUncorMAE}
  \end{subfigure}
  \caption{Comparison of the accuracy of the various corrected and uncorrected schemes for dimer correlation energies}
  \label{fig:dimerCorrectUncorrectStats}
\end{figure}

\begin{figure}[h]
\centering
  \begin{subfigure}{0.49\textwidth}
    \centering
    \includegraphics[width=\textwidth]{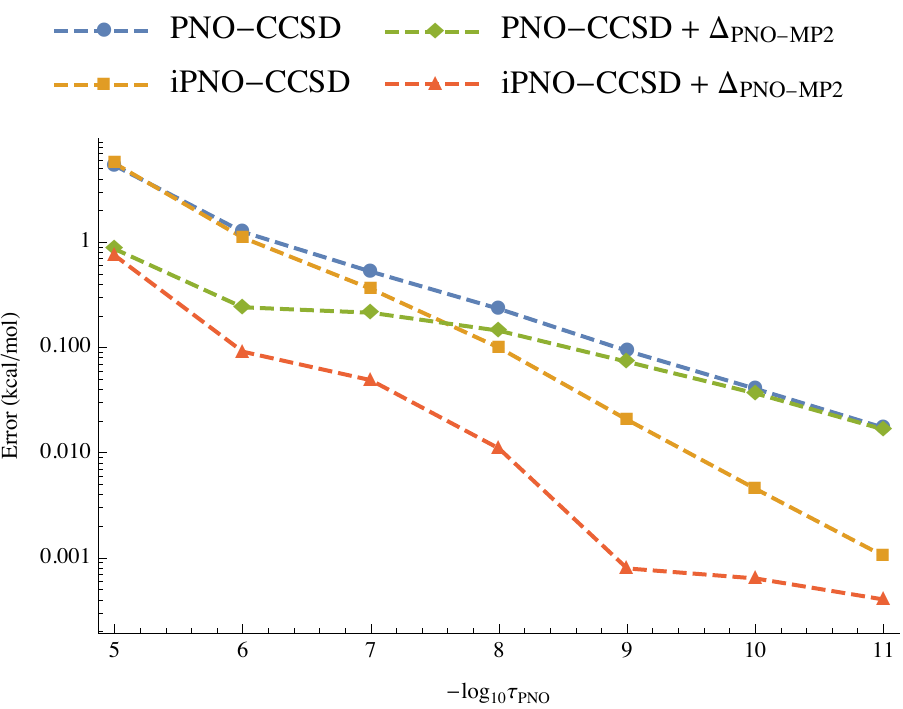}
    \caption{Maximum absolute error (MAX)}
    \label{fig:beCorUncorMAX}
  \end{subfigure}
  \hfill
  \begin{subfigure}{0.49\textwidth}
    \centering
    \includegraphics[width=\textwidth]{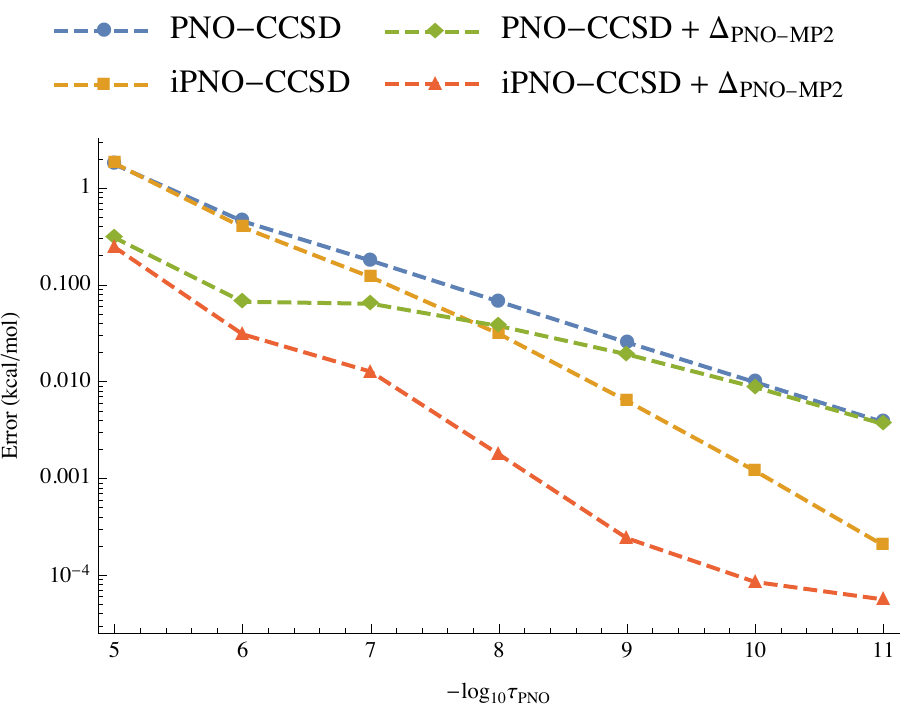}
    \caption{Mean absolute error (MAE)}
    \label{fig:beCorUncorMAE}
  \end{subfigure}
  \caption{Comparison of the accuracy of the various corrected and uncorrected schemes for binding energies}
  \label{fig:beCorrectUncorrectStats}
\end{figure}

Thus the best way to correct the PNO incompleteness of iPNO-CCSD seems to be via $\Delta_\text{PNO-MP2}$. Figures~\ref{fig:dimerCorrectUncorrectStats} and \ref{fig:beCorrectUncorrectStats} compare the performance of PNO-CCSD and iPNO-CCSD with their counterparts corrected with $\Delta_\text{PNO-MP2}$ for correlation energies and binding energies of dimers. For coarse truncations ($\tpno \geq 10^{-7}$), corrected PNO-CCSD energies are more precise than bare (uncorrected) iPNO-CCSD energies.
However, for tighter truncations ($\tpno \leq 10^{-8}$), even uncorrected iPNO-CCSD outperforms corrected PNO-CCSD; this clearly suggests the dominant effect of suboptimal PNOs in the residual error of PNO-CCSD energy in the asymptotic regime ($\tpno \to 0$). Most importantly, note that for $\tpno \leq 10^{-6}$, corrected iPNO-CCSD energies are more precise than both corrected PNO-CCSD and uncorrected iPNO-CCSD. In fact, the performance of corrected iPNO-CCSD is rather remarkable, e.g. max error in binding energy of less than 0.1 kcal/mol is obtained already with $\tpno = 10^{-6}$!


\begin{figure}[h]
\centering
\includegraphics[width=\textwidth]{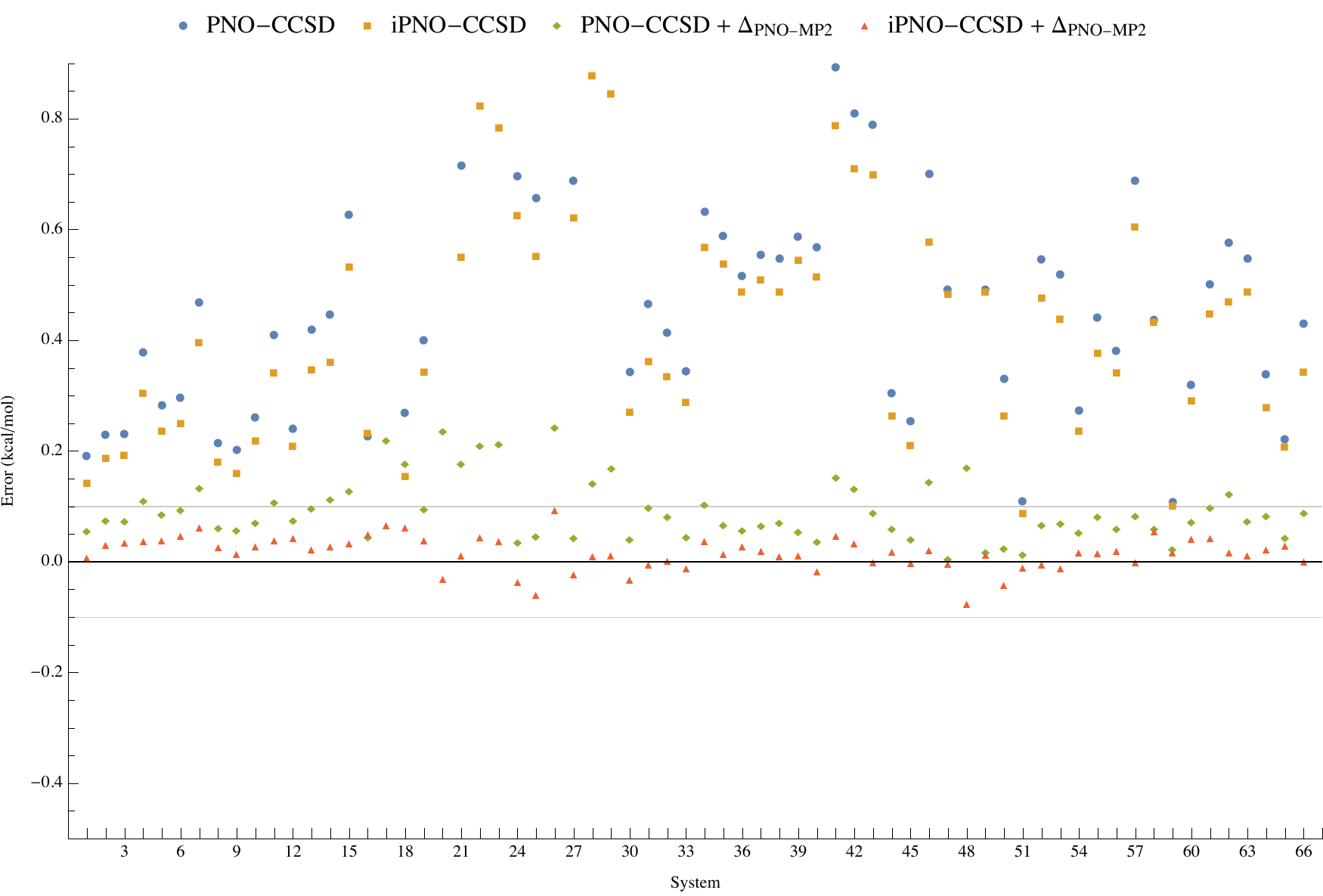}
\caption{Comparison of the binding energy errors for all systems in the S66 benchmark set,
using both corrected and uncorrected PNO-CCSD and iPNO-CCSD schemes}
\label{fig:allS66BEscatter}
\end{figure}

As a further illustration of the performance of our corrected iPNO-CCSD scheme, we
have computed the binding energies for all 66 dimers in the S66 dataset, using both the PNO-CCSD
and iPNO-CCSD schemes with $\tpno = 10^{-6}$. The binding energy errors for the corrected
and uncorrected schemes are plotted in Figure~\ref{fig:allS66BEscatter}, while
Table~\ref{allS66stats} contains a summary of the statistical analysis of these
errors. Of the 66 dimer, not a single one has an
$\text{iPNO-CCSD} + \Delta_{\text{PNO}-\text{MP2}}$ error above 0.1 kcal/mol,
while the mean absolute error for this scheme is more than an order of magnitude
smaller than the corresponding values for uncorrected PNO-CCSD and iPNO-CCSD. Also, the mean absolute error is reduced relative
to the standard $\text{PNO-CCSD} + \Delta_{\text{PNO}-\text{MP2}}$ approach by more than a factor of 3.

\begin{table}
\begin{center}
\begin{tabular}{ccc}
\hline\hline
Method & MAX & MAE \\
\hline
$\text{PNO-CCSD}$ & 4.224 & 0.588 \\
$\text{iPNO-CCSD}$ & 3.958 & 0.514 \\
$\text{PNO-CCSD} + \Delta_{\text{PNO}-\text{MP2}}$ & 0.240 & 0.089 \\
$\text{iPNO-CCSD} + \Delta_{\text{PNO}-\text{MP2}}$ & 0.091 & 0.027 \\
$\text{iPNO-CCSD} + \Delta_{\text{iPNO}-\text{MP2}}$ & 0.756 & 0.112 \\
\hline\hline
\end{tabular}
\caption{Statistical analysis of the binding energy errors (in kcal/mol)
for all S66 systems. In all cases, $\tpno = 10^{-6}$ was used.}
\label{allS66stats}
\end{center}
\end{table}

\section{\label{conclusions}Summary and Perspective}
We have investigated the use of iteratively-optimized PNOs (``iPNOs'') for a more robust compression of the coupled-cluster wave operator in the context of CCSD.
The performance is compared to that of the standard PNO approach in which MP1 PNOs are used to compress the CC wave operator.
PNO optimization offers moderate improvement relative to the standard PNO-CCSD for small PNO ranks but rapidly increases in effectiveness for large PNO ranks;
the PNO incompleteness error of the CCSD energy is reduced by orders of magnitude in the asymptotic regime, with an insignificant increase in PNO ranks.
This suggests that, in applications that call for a precise representation of the wave operator, such as nonresonant response, the use of iPNO-CCSD may be warranted.

The effect of PNO optimization is not accounted for by Neese's perturbative correction for the PNO incompleteness of the CCSD energy. In fact, in the asymptotic regime, the correction becomes ineffective, and the PNO truncation error of PNO-CCSD is entirely dominated by the use of suboptimal (MP1) PNOs. The use of the perturbative correction
in combination with the PNO optimization procedure seems to produce the most precise approximation to the canonical CCSD result for small and large PNO ranks; remarkable improvements with respect to standard PNO approach range from a factor of 3 with $\tpno=10^{-6}$ to more than 2 orders of magnitude with $\tpno=10^{-9}$. Thus, the use of iPNO approach seems warranted even for computing energies. The remarkable performance of perturbatively-corrected iPNO approach for the energies suggests that further investigation of how to correct the PNO incompleteness of the coupled-cluster wave operator (not just the energy) is worthwhile.

The current pilot implementation of the PNO optimization procedure has much room for improvement. The use of the CCSD residual in the full space (or, in the case of domain-based methods, in the full domain space of each pair) is not necessary and can be eliminated by switching to the gradient-based optimization. Gradient-based optimization of PNOs is already used by one of us (E.V.) in the context of real-space correlation methods and seems to be a robust way to compute optimal PNOs without the need to represent the
wave operator explicitly.

\section{Acknowledgments}
This work was partly supported by the Scientific Discovery through
Advanced Computing (SciDAC) program funded by the U.S.~Department of
Energy, Office of Science, Advanced Scientific Computing Research
and Basic Energy Sciences (J. Z. and C. Y.);
by the Exascale Computing Project (ECP), Project Number: 17-SC-20-SC, 
a collaborative effort of two DOE organizations: the Office of Science
and the National Nuclear Security Administration (E. V. and C. Y.);
and by the Virginia Tech Institute for Critical Technology and
Applied Science (ICTAS) (M. C.).
The work at Virginia Tech has also been supported by the U.S. National Science Foundation (awards 1362655 and 1450262).
The authors thank the National Energy Research Scientific Computing 
(NERSC) center and Advanced Research Computing (ARC) at Virginia Tech
for making computational resources available. 

\bibliographystyle{jcc.bst}
\bibliography{refs.bib}

\providecommand{\latin}[1]{#1}
\makeatletter
\providecommand{\doi}
  {\begingroup\let\do\@makeother\dospecials
  \catcode`\{=1 \catcode`\}=2 \doi@aux}
\providecommand{\doi@aux}[1]{\endgroup\texttt{#1}}
\makeatother
\providecommand*\mcitethebibliography{\thebibliography}
\csname @ifundefined\endcsname{endmcitethebibliography}
  {\let\endmcitethebibliography\endthebibliography}{}
\begin{mcitethebibliography}{45}
\providecommand*\natexlab[1]{#1}
\providecommand*\mciteSetBstSublistMode[1]{}
\providecommand*\mciteSetBstMaxWidthForm[2]{}
\providecommand*\mciteBstWouldAddEndPuncttrue
  {\def\EndOfBibitem{\unskip.}}
\providecommand*\mciteBstWouldAddEndPunctfalse
  {\let\EndOfBibitem\relax}
\providecommand*\mciteSetBstMidEndSepPunct[3]{}
\providecommand*\mciteSetBstSublistLabelBeginEnd[3]{}
\providecommand*\EndOfBibitem{}
\mciteSetBstSublistMode{f}
\mciteSetBstMaxWidthForm{subitem}{(\alph{mcitesubitemcount})}
\mciteSetBstSublistLabelBeginEnd
  {\mcitemaxwidthsubitemform\space}
  {\relax}
  {\relax}

\bibitem[Bartlett and Musia{\l}(2007)Bartlett, and Musia{\l}]{Bartlett:2007CC}
Bartlett,~R.; Musia{\l},~M. {Coupled-cluster theory in quantum chemistry}.
  \emph{Rev. Mod. Phys.} \textbf{2007}, \emph{79}, 291\relax
\mciteBstWouldAddEndPuncttrue
\mciteSetBstMidEndSepPunct{\mcitedefaultmidpunct}
{\mcitedefaultendpunct}{\mcitedefaultseppunct}\relax
\EndOfBibitem
\bibitem[Tajti \latin{et~al.}(2004)Tajti, Szalay, Cs{\'{a}}sz{\'{a}}r,
  K{\'{a}}llay, Gauss, Valeev, Flowers, V{\'{a}}zquez, and Stanton]{Tajti2004}
Tajti,~A.; Szalay,~P.~G.; Cs{\'{a}}sz{\'{a}}r,~A.~G.; K{\'{a}}llay,~M.;
  Gauss,~J.; Valeev,~E.~F.; Flowers,~B.~A.; V{\'{a}}zquez,~J.; Stanton,~J.~F.
  {HEAT: High accuracy extrapolated ab initio thermochemistry}. \emph{J. Chem.
  Phys.} \textbf{2004}, \emph{121}, 11599--11613\relax
\mciteBstWouldAddEndPuncttrue
\mciteSetBstMidEndSepPunct{\mcitedefaultmidpunct}
{\mcitedefaultendpunct}{\mcitedefaultseppunct}\relax
\EndOfBibitem
\bibitem[Purvis and Bartlett(1982)Purvis, and Bartlett]{Purvis1982}
Purvis,~G.~D.; Bartlett,~R.~J. \emph{J. Chem. Phys.} \textbf{1982}, \emph{76},
  1910--1918\relax
\mciteBstWouldAddEndPuncttrue
\mciteSetBstMidEndSepPunct{\mcitedefaultmidpunct}
{\mcitedefaultendpunct}{\mcitedefaultseppunct}\relax
\EndOfBibitem
\bibitem[Raghavachari \latin{et~al.}(1989)Raghavachari, Trucks, Pople, and
  Head-Gordon]{Raghavachari:CCSD(T)}
Raghavachari,~K.; Trucks,~G.~W.; Pople,~J.~A.; Head-Gordon,~M. \emph{Chemical
  Physics Letters} \textbf{1989}, \emph{157}, 479--483\relax
\mciteBstWouldAddEndPuncttrue
\mciteSetBstMidEndSepPunct{\mcitedefaultmidpunct}
{\mcitedefaultendpunct}{\mcitedefaultseppunct}\relax
\EndOfBibitem
\bibitem[Bischoff \latin{et~al.}(2012)Bischoff, Harrison, and
  Valeev]{Bischoff:2012fw}
Bischoff,~F.~A.; Harrison,~R.~J.; Valeev,~E.~F. {Computing many-body wave
  functions with guaranteed precision: the first-order M{\o}ller-Plesset wave
  function for the ground state of helium atom.} \emph{J Chem Phys}
  \textbf{2012}, \emph{137}, 104103\relax
\mciteBstWouldAddEndPuncttrue
\mciteSetBstMidEndSepPunct{\mcitedefaultmidpunct}
{\mcitedefaultendpunct}{\mcitedefaultseppunct}\relax
\EndOfBibitem
\bibitem[Bischoff and Valeev(2013)Bischoff, and Valeev]{Bischoff:2013cx}
Bischoff,~F.~A.; Valeev,~E.~F. {Computing molecular correlation energies with
  guaranteed precision}. \emph{J Chem Phys} \textbf{2013}, \emph{139},
  114106\relax
\mciteBstWouldAddEndPuncttrue
\mciteSetBstMidEndSepPunct{\mcitedefaultmidpunct}
{\mcitedefaultendpunct}{\mcitedefaultseppunct}\relax
\EndOfBibitem
\bibitem[Sch{\"a}fer \latin{et~al.}(2017)Sch{\"a}fer, Ramberger, and
  Kresse]{Schafer:2017kq}
Sch{\"a}fer,~T.; Ramberger,~B.; Kresse,~G. {Quartic scaling MP2 for solids: A
  highly parallelized algorithm in the plane wave basis}. \emph{J Chem Phys}
  \textbf{2017}, \emph{146}, 104101\relax
\mciteBstWouldAddEndPuncttrue
\mciteSetBstMidEndSepPunct{\mcitedefaultmidpunct}
{\mcitedefaultendpunct}{\mcitedefaultseppunct}\relax
\EndOfBibitem
\bibitem[Mardirossian \latin{et~al.}(2018)Mardirossian, McClain, and
  Chan]{Mardirossian:2018fk}
Mardirossian,~N.; McClain,~J.~D.; Chan,~G. K.-L. {Lowering of the complexity of
  quantum chemistry methods by choice of representation}. \emph{J Chem Phys}
  \textbf{2018}, \emph{148}, 044106\relax
\mciteBstWouldAddEndPuncttrue
\mciteSetBstMidEndSepPunct{\mcitedefaultmidpunct}
{\mcitedefaultendpunct}{\mcitedefaultseppunct}\relax
\EndOfBibitem
\bibitem[Pinski \latin{et~al.}(2015)Pinski, Riplinger, Valeev, and
  Neese]{Pinski:2015ii}
Pinski,~P.; Riplinger,~C.; Valeev,~E.~F.; Neese,~F. {Sparse maps---A systematic
  infrastructure for reduced-scaling electronic structure methods. I. An
  efficient and simple linear scaling local MP2 method that uses an
  intermediate basis of pair natural orbitals}. \emph{J Chem Phys}
  \textbf{2015}, \emph{143}, 034108\relax
\mciteBstWouldAddEndPuncttrue
\mciteSetBstMidEndSepPunct{\mcitedefaultmidpunct}
{\mcitedefaultendpunct}{\mcitedefaultseppunct}\relax
\EndOfBibitem
\bibitem[Riplinger \latin{et~al.}(2016)Riplinger, Pinski, Becker, Valeev, and
  Neese]{Riplinger:2016dq}
Riplinger,~C.; Pinski,~P.; Becker,~U.; Valeev,~E.~F.; Neese,~F. {Sparse
  maps---A systematic infrastructure for reduced-scaling electronic structure
  methods. II. Linear scaling domain based pair natural orbital coupled cluster
  theory}. \emph{J Chem Phys} \textbf{2016}, \emph{144}, 024109\relax
\mciteBstWouldAddEndPuncttrue
\mciteSetBstMidEndSepPunct{\mcitedefaultmidpunct}
{\mcitedefaultendpunct}{\mcitedefaultseppunct}\relax
\EndOfBibitem
\bibitem[Pavo{\v s}evi{\'c} \latin{et~al.}(2017)Pavo{\v s}evi{\'c}, Peng,
  Pinski, Riplinger, Neese, and Valeev]{Pavosevic:2017kb}
Pavo{\v s}evi{\'c},~F.; Peng,~C.; Pinski,~P.; Riplinger,~C.; Neese,~F.;
  Valeev,~E.~F. {SparseMaps---A systematic infrastructure for reduced scaling
  electronic structure methods. V. Linear scaling explicitly correlated
  coupled-cluster method with pair natural orbitals}. \emph{J Chem Phys}
  \textbf{2017}, \emph{146}, 174108\relax
\mciteBstWouldAddEndPuncttrue
\mciteSetBstMidEndSepPunct{\mcitedefaultmidpunct}
{\mcitedefaultendpunct}{\mcitedefaultseppunct}\relax
\EndOfBibitem
\bibitem[Schwilk \latin{et~al.}(2017)Schwilk, Ma, K{\"o}ppl, and
  Werner]{Schwilk:2017ut}
Schwilk,~M.; Ma,~Q.; K{\"o}ppl,~C.; Werner,~H.-J. {Scalable Electron
  Correlation Methods. 3. Efficient and Accurate Parallel Local Coupled Cluster
  with Pair Natural Orbitals (PNO-LCCSD)}. \emph{J. Chem. Theory Comput.}
  \textbf{2017}, \emph{13}, 3650--3675\relax
\mciteBstWouldAddEndPuncttrue
\mciteSetBstMidEndSepPunct{\mcitedefaultmidpunct}
{\mcitedefaultendpunct}{\mcitedefaultseppunct}\relax
\EndOfBibitem
\bibitem[Ma \latin{et~al.}(2017)Ma, Schwilk, K{\"o}ppl, and Werner]{Ma:2017ef}
Ma,~Q.; Schwilk,~M.; K{\"o}ppl,~C.; Werner,~H.-J. {Scalable Electron
  Correlation Methods. 4. Parallel Explicitly Correlated Local Coupled Cluster
  with Pair Natural Orbitals (PNO-LCCSD-F12)}. \emph{J. Chem. Theory Comput.}
  \textbf{2017}, \emph{13}, 4871--4896\relax
\mciteBstWouldAddEndPuncttrue
\mciteSetBstMidEndSepPunct{\mcitedefaultmidpunct}
{\mcitedefaultendpunct}{\mcitedefaultseppunct}\relax
\EndOfBibitem
\bibitem[Ma and Werner(2018)Ma, and Werner]{Ma:2018fl}
Ma,~Q.; Werner,~H.-J. {Scalable Electron Correlation Methods. 5. Parallel
  Perturbative Triples Correction for Explicitly Correlated Local Coupled
  Cluster with Pair Natural Orbitals}. \emph{J. Chem. Theory Comput.}
  \textbf{2018}, \emph{14}, 198--215\relax
\mciteBstWouldAddEndPuncttrue
\mciteSetBstMidEndSepPunct{\mcitedefaultmidpunct}
{\mcitedefaultendpunct}{\mcitedefaultseppunct}\relax
\EndOfBibitem
\bibitem[Schmitz \latin{et~al.}(2014)Schmitz, H{\"a}ttig, and
  Tew]{Schmitz:2014do}
Schmitz,~G.; H{\"a}ttig,~C.; Tew,~D.~P. {Explicitly correlated PNO-MP2 and
  PNO-CCSD and their application to the S66 set and large molecular systems}.
  \emph{Phys Chem Chem Phys} \textbf{2014}, \emph{16}, 22167--22178\relax
\mciteBstWouldAddEndPuncttrue
\mciteSetBstMidEndSepPunct{\mcitedefaultmidpunct}
{\mcitedefaultendpunct}{\mcitedefaultseppunct}\relax
\EndOfBibitem
\bibitem[Schmitz and H{\"a}ttig(2016)Schmitz, and H{\"a}ttig]{Schmitz:2016bu}
Schmitz,~G.; H{\"a}ttig,~C. {Perturbative triples correction for local pair
  natural orbital based explicitly correlated CCSD(F12*) using Laplace
  transformation techniques}. \emph{J Chem Phys} \textbf{2016}, \emph{145},
  234107\relax
\mciteBstWouldAddEndPuncttrue
\mciteSetBstMidEndSepPunct{\mcitedefaultmidpunct}
{\mcitedefaultendpunct}{\mcitedefaultseppunct}\relax
\EndOfBibitem
\bibitem[Neese \latin{et~al.}(2009)Neese, Wennmohs, and
  Hansen]{Neese:2009_130_114108}
Neese,~F.; Wennmohs,~F.; Hansen,~A. {Efficient and accurate local
  approximations to coupled-electron pair approaches: An attempt to revive the
  pair natural orbital method}. \emph{J. Chem. Phys.} \textbf{2009},
  \emph{130}, 114108\relax
\mciteBstWouldAddEndPuncttrue
\mciteSetBstMidEndSepPunct{\mcitedefaultmidpunct}
{\mcitedefaultendpunct}{\mcitedefaultseppunct}\relax
\EndOfBibitem
\bibitem[Neese \latin{et~al.}(2009)Neese, Hansen, and Liakos]{Neese2009}
Neese,~F.; Hansen,~A.; Liakos,~D.~G. {Efficient and accurate approximations to
  the local coupled cluster singles doubles method using a truncated pair
  natural orbital basis}. \emph{J. Chem. Phys.} \textbf{2009}, \emph{131},
  064103\relax
\mciteBstWouldAddEndPuncttrue
\mciteSetBstMidEndSepPunct{\mcitedefaultmidpunct}
{\mcitedefaultendpunct}{\mcitedefaultseppunct}\relax
\EndOfBibitem
\bibitem[Yang \latin{et~al.}(2011)Yang, Kurashige, Manby, and
  Chan]{Yang:2011_OSV}
Yang,~J.; Kurashige,~Y.; Manby,~F.~R.; Chan,~G. K.~L. \emph{J. Chem. Phys.}
  \textbf{2011}, \emph{134}, 044123\relax
\mciteBstWouldAddEndPuncttrue
\mciteSetBstMidEndSepPunct{\mcitedefaultmidpunct}
{\mcitedefaultendpunct}{\mcitedefaultseppunct}\relax
\EndOfBibitem
\bibitem[Yang \latin{et~al.}(2012)Yang, Chan, Manby, Sch\"{u}tz, and
  Werner]{Yang:2012_OSV}
Yang,~J.; Chan,~G. K.~L.; Manby,~F.~R.; Sch\"{u}tz,~M.; Werner,~H.-J. \emph{J.
  Chem. Phys.} \textbf{2012}, \emph{136}, 144105\relax
\mciteBstWouldAddEndPuncttrue
\mciteSetBstMidEndSepPunct{\mcitedefaultmidpunct}
{\mcitedefaultendpunct}{\mcitedefaultseppunct}\relax
\EndOfBibitem
\bibitem[Riplinger \latin{et~al.}(2013)Riplinger, Sandhoefer, Hansen, and
  Neese]{Riplinger:2013_TriplesNOs}
Riplinger,~C.; Sandhoefer,~B.; Hansen,~A.; Neese,~F. \emph{J. Chem. Phys.}
  \textbf{2013}, \emph{139}, 134101\relax
\mciteBstWouldAddEndPuncttrue
\mciteSetBstMidEndSepPunct{\mcitedefaultmidpunct}
{\mcitedefaultendpunct}{\mcitedefaultseppunct}\relax
\EndOfBibitem
\bibitem[L{\"{o}}wdin(1955)]{Lowdin1955}
L{\"{o}}wdin,~P.-O. \emph{Physical Review} \textbf{1955}, \emph{97},
  1474--1489\relax
\mciteBstWouldAddEndPuncttrue
\mciteSetBstMidEndSepPunct{\mcitedefaultmidpunct}
{\mcitedefaultendpunct}{\mcitedefaultseppunct}\relax
\EndOfBibitem
\bibitem[Edmiston and Krauss(1966)Edmiston, and Krauss]{Edmiston:1966_PNO}
Edmiston,~C.; Krauss,~M. Pseudonatural Orbitals as a Basis for the
  Superposition of Configurations. I. He2+. \emph{J. Chem. Phys.}
  \textbf{1966}, \emph{45}, 1833\relax
\mciteBstWouldAddEndPuncttrue
\mciteSetBstMidEndSepPunct{\mcitedefaultmidpunct}
{\mcitedefaultendpunct}{\mcitedefaultseppunct}\relax
\EndOfBibitem
\bibitem[Edmiston and Krauss(1968)Edmiston, and Krauss]{Edmiston:1968_PNO}
Edmiston,~C.; Krauss,~M. \emph{J. Chem. Phys.} \textbf{1968}, \emph{49},
  192--205\relax
\mciteBstWouldAddEndPuncttrue
\mciteSetBstMidEndSepPunct{\mcitedefaultmidpunct}
{\mcitedefaultendpunct}{\mcitedefaultseppunct}\relax
\EndOfBibitem
\bibitem[Meyer(1971)]{Meyer:1971_PNO}
Meyer,~W. \emph{International Journal of Quantum Chemistry} \textbf{1971},
  \emph{5}, 341--348\relax
\mciteBstWouldAddEndPuncttrue
\mciteSetBstMidEndSepPunct{\mcitedefaultmidpunct}
{\mcitedefaultendpunct}{\mcitedefaultseppunct}\relax
\EndOfBibitem
\bibitem[Meyer(1973)]{Meyer:1973_PNO}
Meyer,~W. \emph{J. Chem. Phys.} \textbf{1973}, \emph{58}, 1017--1035\relax
\mciteBstWouldAddEndPuncttrue
\mciteSetBstMidEndSepPunct{\mcitedefaultmidpunct}
{\mcitedefaultendpunct}{\mcitedefaultseppunct}\relax
\EndOfBibitem
\bibitem[Meyer and Rosmus(1975)Meyer, and Rosmus]{Meyer:1975_PNO}
Meyer,~W.; Rosmus,~P. \emph{J. Chem. Phys.} \textbf{1975}, \emph{63},
  2356\relax
\mciteBstWouldAddEndPuncttrue
\mciteSetBstMidEndSepPunct{\mcitedefaultmidpunct}
{\mcitedefaultendpunct}{\mcitedefaultseppunct}\relax
\EndOfBibitem
\bibitem[Ahlrichs \latin{et~al.}(1975)Ahlrichs, Lischka, Staemmler, and
  Kutzelnigg]{Ahlrichs1975}
Ahlrichs,~R.; Lischka,~H.; Staemmler,~V.; Kutzelnigg,~W. \emph{J. Chem. Phys.}
  \textbf{1975}, \emph{62}, 1225--1234\relax
\mciteBstWouldAddEndPuncttrue
\mciteSetBstMidEndSepPunct{\mcitedefaultmidpunct}
{\mcitedefaultendpunct}{\mcitedefaultseppunct}\relax
\EndOfBibitem
\bibitem[Sch\"{u}tz and Werner(2001)Sch\"{u}tz, and Werner]{Schutz:2001_LCCSD}
Sch\"{u}tz,~M.; Werner,~H.-J. \emph{J. Chem. Phys.} \textbf{2001}, \emph{114},
  661\relax
\mciteBstWouldAddEndPuncttrue
\mciteSetBstMidEndSepPunct{\mcitedefaultmidpunct}
{\mcitedefaultendpunct}{\mcitedefaultseppunct}\relax
\EndOfBibitem
\bibitem[Sch{\"{u}}tz(2002)]{Schutz:2002}
Sch{\"{u}}tz,~M. \emph{Phys. Chem. Chem. Phys.} \textbf{2002}, \emph{4},
  3941--3947\relax
\mciteBstWouldAddEndPuncttrue
\mciteSetBstMidEndSepPunct{\mcitedefaultmidpunct}
{\mcitedefaultendpunct}{\mcitedefaultseppunct}\relax
\EndOfBibitem
\bibitem[Sch{\"{u}}tz and Manby(2003)Sch{\"{u}}tz, and Manby]{Schutz:2003}
Sch{\"{u}}tz,~M.; Manby,~F.~R. \emph{Phys. Chem. Chem. Phys.} \textbf{2003},
  \emph{5}, 3349--3358\relax
\mciteBstWouldAddEndPuncttrue
\mciteSetBstMidEndSepPunct{\mcitedefaultmidpunct}
{\mcitedefaultendpunct}{\mcitedefaultseppunct}\relax
\EndOfBibitem
\bibitem[Riplinger and Neese(2013)Riplinger, and Neese]{Riplinger:2013ek}
Riplinger,~C.; Neese,~F. {An efficient and near linear scaling pair natural
  orbital based local coupled cluster method.} \emph{J Chem Phys}
  \textbf{2013}, \emph{138}, 034106\relax
\mciteBstWouldAddEndPuncttrue
\mciteSetBstMidEndSepPunct{\mcitedefaultmidpunct}
{\mcitedefaultendpunct}{\mcitedefaultseppunct}\relax
\EndOfBibitem
\bibitem[Riplinger \latin{et~al.}(2013)Riplinger, Sandhoefer, Hansen, and
  Neese]{Riplinger:2013tn}
Riplinger,~C.; Sandhoefer,~B.; Hansen,~A.; Neese,~F. {Natural triple
  excitations in local coupled cluster calculations with pair natural
  orbitals.} \emph{J Chem Phys} \textbf{2013}, \emph{139}, 134101\relax
\mciteBstWouldAddEndPuncttrue
\mciteSetBstMidEndSepPunct{\mcitedefaultmidpunct}
{\mcitedefaultendpunct}{\mcitedefaultseppunct}\relax
\EndOfBibitem
\bibitem[Pavo{\v s}evi{\'c} \latin{et~al.}(2016)Pavo{\v s}evi{\'c}, Pinski,
  Riplinger, Neese, and Valeev]{Pavosevic:2016bc}
Pavo{\v s}evi{\'c},~F.; Pinski,~P.; Riplinger,~C.; Neese,~F.; Valeev,~E.~F.
  {SparseMaps---A systematic infrastructure for reduced-scaling electronic
  structure methods. IV. Linear-scaling second-order explicitly correlated
  energy with pair natural orbitals}. \emph{J Chem Phys} \textbf{2016},
  \emph{144}, 144109\relax
\mciteBstWouldAddEndPuncttrue
\mciteSetBstMidEndSepPunct{\mcitedefaultmidpunct}
{\mcitedefaultendpunct}{\mcitedefaultseppunct}\relax
\EndOfBibitem
\bibitem[Saitow \latin{et~al.}(2017)Saitow, Becker, Riplinger, Valeev, and
  Neese]{Saitow:2017bo}
Saitow,~M.; Becker,~U.; Riplinger,~C.; Valeev,~E.~F.; Neese,~F. {A new
  near-linear scaling, efficient and accurate, open-shell domain-based local
  pair natural orbital coupled cluster singles and doubles theory}. \emph{J
  Chem Phys} \textbf{2017}, \emph{146}, 164105\relax
\mciteBstWouldAddEndPuncttrue
\mciteSetBstMidEndSepPunct{\mcitedefaultmidpunct}
{\mcitedefaultendpunct}{\mcitedefaultseppunct}\relax
\EndOfBibitem
\bibitem[Tew \latin{et~al.}(2011)Tew, Helmich, and H{\"{a}}ttig]{Tew:2011}
Tew,~D.~P.; Helmich,~B.; H{\"{a}}ttig,~C. \emph{J. Chem. Phys.} \textbf{2011},
  \emph{135}, 074107\relax
\mciteBstWouldAddEndPuncttrue
\mciteSetBstMidEndSepPunct{\mcitedefaultmidpunct}
{\mcitedefaultendpunct}{\mcitedefaultseppunct}\relax
\EndOfBibitem
\bibitem[Korona and Werner(2003)Korona, and Werner]{Korona:2003_simulation}
Korona,~T.; Werner,~H.-J. \emph{J. Chem. Phys.} \textbf{2003}, \emph{118},
  3006--3019\relax
\mciteBstWouldAddEndPuncttrue
\mciteSetBstMidEndSepPunct{\mcitedefaultmidpunct}
{\mcitedefaultendpunct}{\mcitedefaultseppunct}\relax
\EndOfBibitem
\bibitem[Krause and Werner(2012)Krause, and Werner]{Krause:2012_simulation}
Krause,~C.; Werner,~H.-J. \emph{Physical Chemistry Chemical Physics}
  \textbf{2012}, \emph{14}, 7591\relax
\mciteBstWouldAddEndPuncttrue
\mciteSetBstMidEndSepPunct{\mcitedefaultmidpunct}
{\mcitedefaultendpunct}{\mcitedefaultseppunct}\relax
\EndOfBibitem
\bibitem[Peng \latin{et~al.}()Peng, Clement, and
  Valeev]{Peng:2018_PNO-EOM-CCSD}
Peng,~C.; Clement,~M.~C.; Valeev,~E.~F. arXiv:1802.06738\relax
\mciteBstWouldAddEndPuncttrue
\mciteSetBstMidEndSepPunct{\mcitedefaultmidpunct}
{\mcitedefaultendpunct}{\mcitedefaultseppunct}\relax
\EndOfBibitem
\bibitem[Valeev \latin{et~al.}()Valeev, Peng, Lewis, and Calvin]{MPQC4}
Valeev,~E.~F.; Peng,~C.; Lewis,~C.~A.; Calvin,~J.~A. The Massively Parallel
  Quantum chemistry Program (MPQC), Version 4.0.0.
  \url{https://github.com/ValeevGroup/mpqc4}\relax
\mciteBstWouldAddEndPuncttrue
\mciteSetBstMidEndSepPunct{\mcitedefaultmidpunct}
{\mcitedefaultendpunct}{\mcitedefaultseppunct}\relax
\EndOfBibitem
\bibitem[{\v R}ez{\`a}{\v c}˜ \latin{et~al.}(2011){\v R}ez{\`a}{\v c}˜, Riley,
  and Hobza]{Hobza:2011_S66}
{\v R}ez{\`a}{\v c}˜,~J.; Riley,~K.~E.; Hobza,~P. S66: A Well-balanced Database
  of Benchmark Interaction Energies Relevant to Biomolecular Structures.
  \emph{J. Chem. Theory Comput.} \textbf{2011}, \emph{7}, 2427\relax
\mciteBstWouldAddEndPuncttrue
\mciteSetBstMidEndSepPunct{\mcitedefaultmidpunct}
{\mcitedefaultendpunct}{\mcitedefaultseppunct}\relax
\EndOfBibitem
\bibitem[{\v R}ez{\`a}{\v c}˜ \latin{et~al.}(2008){\v R}ez{\`a}{\v c}˜, Jure{\v
  c}˜ka, Riley, {\v C}ern{\' y}, Valdes, Pluh{\' a}{\v c}kov{\' a}, Berka, {\v
  R}ez{\' a}{\v c}, Pito{\v n}{\' a}k, Vondr{\' a}{\v s}ek, and Hobza]{begdb}
{\v R}ez{\`a}{\v c}˜,~J.; Jure{\v c}˜ka,~P.; Riley,~K.~E.; {\v C}ern{\' y},~J.;
  Valdes,~H.; Pluh{\' a}{\v c}kov{\' a},~K.; Berka,~K.; {\v R}ez{\' a}{\v
  c},~T.; Pito{\v n}{\' a}k,~M.; Vondr{\' a}{\v s}ek,~J.; Hobza,~P. Quantum
  Chemical Benchmark Energy and Geometry Database for Molecular Clusters and
  Complex Molecular Systems (www.begdb.com): A Users Manual and Examples.
  \emph{Collect. Czech. Chem. Commun.} \textbf{2008}, \emph{73}, 1261\relax
\mciteBstWouldAddEndPuncttrue
\mciteSetBstMidEndSepPunct{\mcitedefaultmidpunct}
{\mcitedefaultendpunct}{\mcitedefaultseppunct}\relax
\EndOfBibitem
\bibitem[Peterson \latin{et~al.}(2008)Peterson, Adler, and Werner]{ccpVDZF12}
Peterson,~K.~A.; Adler,~T.~B.; Werner,~H.-J. \emph{J. Chem. Phys.}
  \textbf{2008}, \emph{128}, 084102\relax
\mciteBstWouldAddEndPuncttrue
\mciteSetBstMidEndSepPunct{\mcitedefaultmidpunct}
{\mcitedefaultendpunct}{\mcitedefaultseppunct}\relax
\EndOfBibitem
\bibitem[Weigend \latin{et~al.}(2002)Weigend, K\"{o}hn, and
  H\"{a}ttig]{augccpVDZRI}
Weigend,~F.; K\"{o}hn,~A.; H\"{a}ttig,~C. \emph{J. Chem. Phys.} \textbf{2002},
  \emph{116}, 3175--3183\relax
\mciteBstWouldAddEndPuncttrue
\mciteSetBstMidEndSepPunct{\mcitedefaultmidpunct}
{\mcitedefaultendpunct}{\mcitedefaultseppunct}\relax
\EndOfBibitem
\end{mcitethebibliography}

\end{document}